\documentclass[amsmath,aps,prl,twocolumn]{revtex4}
\usepackage{amsthm,amsfonts,graphicx,verbatim,color,cancel,amsmath,amssymb}
\newcommand{\mathsym}[1]{{}}
\newcommand{\unicode}[1]{{}}

\def\d{\delta}

\newcommand{\bea}{\begin{eqnarray}}
\newcommand{\eea}{\end{eqnarray}}
\newcommand{\bra}{\langle}
\newcommand{\ket}{\rangle}

\newcommand{\D}{{\rm d}}
\begin{document}

\title{Entanglement Hamiltonians for chiral fermions with zero modes}

\author{Israel Klich, Diana Vaman and Gabriel Wong}
 \affiliation
    {%
    Department of Physics,
    University of Virginia, Box 400714,
    Charlottesville, Virginia 22904, USA
    }%

\affiliation{Department of Physics, University of Virginia, Charlottesville, VA,
USA}
\begin{abstract}
In this Letter, we study the effect of topological zero modes on Entanglement Hamiltonians and entropy of free chiral fermions in (1+1)d. We show how Riemann-Hilbert solutions combined with finite rank perturbation theory allow us to obtain exact expressions for Entanglement Hamiltonians. In the absence of the zero mode, the resulting Entanglement Hamiltonians consists of  local and bi-local terms.
In the periodic sector, the presence of a zero mode leads to an additional non-local contribution to the entanglement Hamiltonian.  We derive an exact expression for this term and for the resulting  change in  the  entanglement entropy.
\end{abstract}
\maketitle

Entanglement Hamiltonians (EH) are the next object to explore in a series of advances in our understanding the quantum structure of many-body states in field theory and condensed matter \cite{casini2009reduced,peschel2011relation,lauchli2012entanglement,qi2012general,hermanns2014entanglement,poilblanc2014entanglement,cardy2016entanglement,rispler2015long}. Indeed, much work has been devoted to understanding entanglement entropy, in a variety of systems, and, subsequently, more detailed information about the spectrum of reduced density matrices is being investigated as well. Entanglement Hamiltonians go one step further in that they contain information about the entanglement spectrum, as well as about the associated eigenvectors, and the possibility to understand the reduced state as a thermal state. Perhaps the most striking of the recent results about Entanglement Hamiltonians is the realization that for a spherical entangling region in a conformal field theory (CFT), Entanglement Hamiltonians have a local form, that may be interpreted as the original hamiltonian energy density with a spatially dependent temperature \cite{bisognano1976duality,myers,Wong:2013fk}.
This result is unusual in its simplicity, as, unfortunately, the computation of EHs is, in general, much more involved than entropy and spectrum and only a few results are available. Thus it is a grand challenge to find additional solvable cases.  

In this Letter, we present a new method for computing Entanglement Hamiltonians of free fermions in the presence of zero modes. 
Fermionic modes localized on topological defects often reflect the topological nature of the defect through unusual properties such as charge fractionalization  \cite{jackiw1976solitons}  and non-abelian braiding \cite{ivanov2001non} and appear in a variety of systems from polyacetylene  \cite{su1979solitons} to defects in topological insulators and superconductors \cite{read2000paired,hou2007electron,teo2010topological}.
The modes are usually intimately tied to the appearance of ground state degeneracies of a topological nature. 
Here, we are able to present new results for the EHs for chiral fermions, and study in detail the effects of the boundary conditions (BC), periodic/anti-periodic for Majorana and generic for Dirac fermions, and of the zero modes (present in the case of periodic boundary conditions) on entanglement. The edge theory of the $p+ ip $ superconductor provides an explicit realization of such a scenario for Majorana fermions and serves as a concrete physical model for our calculation (see, e.g. \cite{fendley2007edge}).
Meanwhile, our calculation in the Dirac sector is relevant to the edge theory of the Quantum Hall state. Wherever available we make contact with previous results in the literature. 

Our approach, as we will soon review, starts with the relation between the EH and the resolvent associated with an appropriately projected (onto the entangling region) free fermion Green's function, and adding the contribution of zero modes with an essentially exact re-summation.  For concreteness we concentrate on chiral Majorana and Dirac fermions in $(1+1)d$, with the spatial direction taken to be a finite radius circle. As we find below, in this case, the computation of the resolvent can be recast into a Riemann-Hilbert problem (RHP, see below), generalizing the approach of Casini and Huerta \cite{casini2009reduced} who computed the EH for free chiral fermions, albeit without zero modes. We combine the RHP approach with exactly summable perturbation theory, to compute the EH for chiral Majorana and Dirac fermions on a finite circle. As our main result we find an exact analytic form for the entanglement Hamiltonian Eq.  \eqref{H Dirac Local}, with a contribution from the topological zero given separately in Eq. \eqref{HzeroMode}.


The reduced density matrix $\rho_{V}$  on a spatial region $V$ is defined to reproduce expectation values of all operators $O_{V}$ inside $V$ via the relation $\rm Tr_{V}(\rho_{V} O_V ) \!=\!\bra O_{V} \ket $. The EH  $\mathcal{H}_V$ is an effective Hamiltonian inside $V$ given by
\bea
 \rho_{V}=\frac{e^{-\mathcal{H}_V}}{Z_V} .
\eea
For free fermions and in the absence of a ground state degeneracy (i.e. in the absence of zero modes), Wick's Theorem implies that the EH is a quadratic operator whose kernel (the single particle EH) $H_{V}$  is uniquely determined by the equal-time Green's function 
\bea\label{greens111}
G(x,y)=\langle \Psi(x) \Psi^{\dag}(y)\rangle
\eea
via the relation 
$H_{V}=-\log((P_VG P_V)^{-1}-1)$ \cite{Peschel}. Here $P_V G P_V$ is the Green's function restricted to the region $V$ by the projectors $P_V$.  As we demonstrate in the supplementary materials, a similar relation holds for Majorana fermions $\psi$, given $\langle\psi(x)\psi(y)\rangle$, allowing us to treat the problem on a parallel footing.

The relation between $H_V$ and $G$ may be expressed in integral form as:
\bea \label{EH}
 {H_{V}}=-\int _{\frac{1}{2}}^{\infty }\left(L(\beta)+L(-\beta)\right)\D\beta,
\eea
where $L(\beta)$ is the resolvent  \cite{casini2009reduced}
\bea\label{resolvent}
L(\beta)=(P_VGP_V - 1/2 +\beta)^{-1}.
\eea
Thus the derivation of the EH reduces to a computation of a projected Green's function resolvent.
It turns out that finding the resolvent $L(\beta)$ can be recast as an RHP. 

This connection between the resolvent and the RHP arises in the following way: for free fermions (and in the absence of zero modes), 
the Green's function $G(x,y)$ computed on the ground state, \emph{un}-restricted to some entangling region $V$,  acts as a single particle operator that projects onto positive energy modes.  For chiral fermions in $(1+1)d$, the projection onto positive modes can be interpreted as a projection onto analytic functions on the upper/lower half of the complex plane. 
For such a $G(x,y)$, the computation of the resolvent \eqref{resolvent} (associated with the projection of $G$ onto $V$) can be mapped onto a RHP (see, e.g. \cite{deift2000orthogonal}), which, in the simple case treated here is a scalar RHP and so exactly solvable. Succinctly put, a  RHP is a jump problem along some smooth contour $\mathcal{C}$ in the complex plane, for a piecewise analytic matrix function. Solving it amounts to finding the matrix function $X(z)$ which is analytic on  $\mathbb{C}\diagdown \mathcal{C}$, and which has boundary values on both sides of $\mathcal{C}$, $X_{\pm}(z)$ subject to a jump condition $X_{-}^{-1}(s)X_+(s)=v(s)$ for $s\in \mathcal{C}$, and with a given matrix $v(s)$. The jump function for our case is a scalar determined by the projector onto $V$, $P_V$.

Our next idea is that once we have access to the resolvent \eqref{resolvent} (either through the RHP or any other method), we can use it to explore various deformations of the system that yield simple  perturbations of the Green's function \eqref{greens111}. In particular, topologically nontrivial background for the fermion will result in the appearance of zero modes \cite{callias1978axial,weinberg1981index}. As we show below, these manifest themselves as {\it finite rank} perturbations of the correlation matrices.   Given the resolvent for the unperturbed Green's function, we solve for the perturbed resolvent by summing up exactly the resulting perturbative series. Here we carry out this recipe for the simplest scenario of a chiral Dirac/Majorana fermion on a spatial circle with general boundary conditions.  

Starting with anti-periodic (Neveu-Schwarz, NS) fermions on a circle with no gauge field, inserting a $\pi$ flux then takes the anti-periodic sector to the periodic (Ramond, R) sector, which has a zero mode whose effect on the EH we derive explicitly.  Note that the appearance of ground state degeneracy for period boundary conditions has been studied extensively. In particular,  Affleck and Ludwig \cite{affleck1991universal}, introduced the notion of boundary entropy, to understand the change in degeneracy due to impurities.  It was also noted in \cite{calabrese2004entanglement} that the boundary entropy can be manifested in a sub-leading contribution to the EE of a finite interval. While these calculations are usually carried out by looking at the entire system, or at EE, our EH calculation allows us to study the effect of degeneracy from the point of view of the local density matrix. In Eq.  \eqref{deltasint}, we find the exact change in the entanglement entropy (EE) of free chiral fermions between periodic and anti-periodic BC, for an arbitrary subset of the circle, and the term in the EH associated with it. In the limiting case when the entangling region spans the circle, this "local boundary entropy" reproduces the boundary entropy. 

 \begin{figure}
\includegraphics[scale=.4]{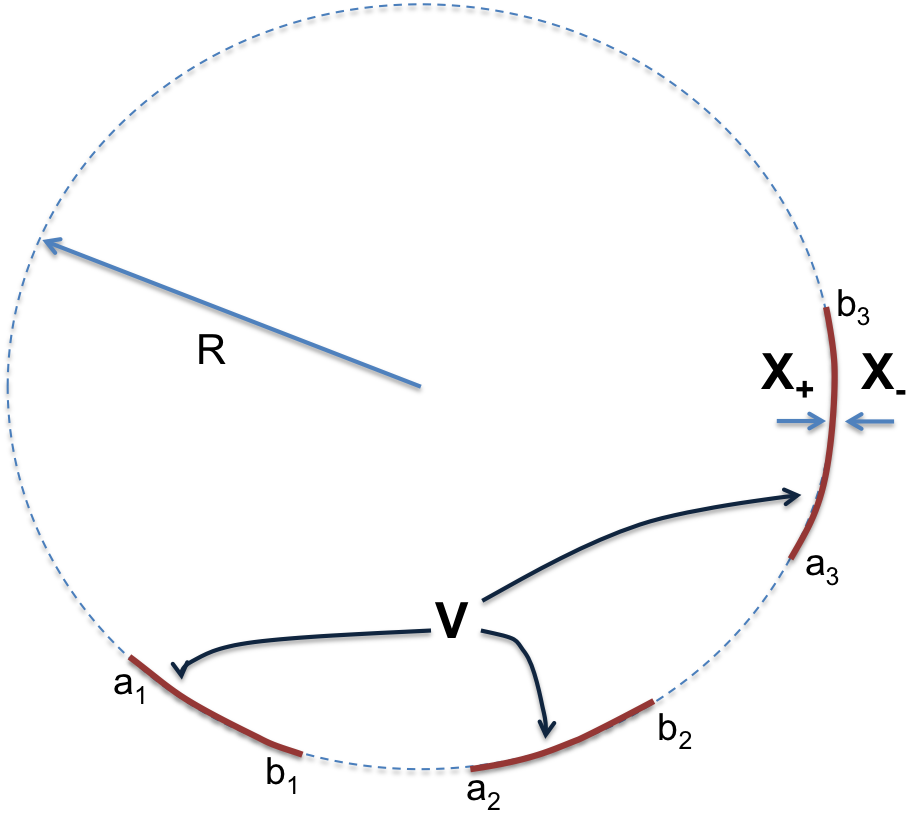}
\caption{The entanglement region $V=\cup( a_{j,}b_j )$ for fermions on a ring of radius $R$ considered in the paper, and an associated Riemann-Hilbert problem.}\label{Entangling region}
\end{figure}
Concretely, we consider the EH on a region $V=\cup( a_{j,}b_j )$ as depicted in Fig. \ref{Entangling region}. Consider a chiral Dirac fermion in on a ring with periodicity $x\!\sim\! x+2\pi R$.  Assuming generic BC,  $\Psi(x+2\pi R)=\Psi(x) e^{2\pi \alpha i}$, the mode expansion is $\Psi(x,t)\!=\!\frac{1}{\sqrt{2\pi R}}\sum_{k\in  {\mathbb{Z}}+\alpha} b_{k}e^{-i \tfrac{k}R (t-x)}$, where $\alpha\in[0,1)$.  
Here $\alpha=\frac{1}{2}$ for the NS sector and  $\alpha=0 $ for the R sector. 
For  $\alpha\neq 0$, evaluating the equal time Green's function on the ground state $|\Omega\rangle$, defined by $b_{k>0}|\Omega\rangle=b^\dagger_{k<0}|\Omega
\rangle=0$,
 gives  
\bea 
G^{ \alpha }(x,y)\!\equiv\! \langle \Omega|\Psi  (x)\Psi^\dagger (y)|\Omega\rangle
\!=\!e^{\frac{i\alpha  (x-y)}{ R}}n(x,y).
\eea
where $n(x,y)\equiv\langle x|n|y\rangle$ is the kernel of the single particle projector $n$ onto momentum modes $\langle x|k\rangle=\frac{1}{\sqrt{2\pi  R}}e^{\frac{ i k x}{ R}}$ with $k$ a non-negative integer \footnote{For our right-moving sector chiral Lagrangian, the single particle momentum states $|k\rangle$ are states of energy $E_{k}=k$. Therefore a projector onto $k\geq 0$ momentum modes is also a projector onto states with $E_{k}\geq 0$.}.
Explicitly:
\bea\label{n projector}
 n(x,y)\equiv {1\over 2\pi R} \sum_{k=0}^{\infty} e^{\frac{i k (x-y+i 0^+)}{R} }
.\eea

We note that the $G^{ \alpha }$ for $\alpha \neq 0$ are related by spectral flow, implemented by a similarity transformation
\bea \label {g}
G^{ \alpha }&=&U_{\alpha }n U_{\alpha }^{-1}, \;\;\; \alpha \neq 0,
\eea
where $U_{\alpha}$ are unitary transformations which translate momenta by $\alpha$: 
$U_{\alpha}|k\rangle=|k+\alpha\rangle$. Moreover,  $U_{\alpha}$ commutes with spatial projection $P_V$, which implies that the projected resolvents (\ref{resolvent}) and EH kernel (\ref{EH}) for different $\alpha$ are also related by a similarity transform. 
Thus, we can relate all the $\alpha \neq 0 $ resolvents $L^{\alpha } = U_{\alpha }N U_{\alpha }^{-1} $ to the following one:
\bea
\label{N beta Res}
N(\beta )\equiv \bigg(P_V n P_V+ \beta  -\tfrac{1}{2}\bigg)^{-1} \label{n0}
\label{Lalpha}  
\eea

As an operator acting on functions defined on $V$, $N(\beta)$ is the resolvent for $P_{V} n P_{V}$, which is an  {\it integrable operator}  in the sense of \cite{its1990differential,deift2000orthogonal}. In general, such operators are defined on a curve $\mathcal{C}$ on a manifold 
 conformally related to the Riemann Sphere and have a kernel of the form 
\bea\label{K}
K(x,y) =\sum_{i} {f}_i(x) {n}(x,y) {g}_i(y),
\eea
where $n(x,y)$ is a projector onto analytic functions  "inside" $\Sigma$, and ${f_{i}},{g_{i}}$ are a set of given functions of $\Sigma$. It is well known that the resolvent, $(1+K)^{-1}$, is related to the solution of a matrix RH problem \cite{deift2000orthogonal}. In our case 
$\mathcal{C}=S^{1}$ on the compactified complex plane Fig. \ref{Entangling region}.  Since we only have a single term in $K$, the matrix RHP reduces to the scalar one and can be stated as follows. 
Define the function $X(x)$ on $S^1$ in terms of the position space elements  $f$ and $g$ by 
$X(x)=1+1/(\beta-\tfrac12)f (x) g(x)$.  The scalar RHP amounts to finding functions $X_{+ }/X_-$ that are analytic inside/outside the unit disc satisfying
\bea\label{RHP}
X(x) =X_{-}^{-1}(x)X_+(x), \;\;\; x \in S^1.
\eea 
Equivalently, we can characterize $X_{+ }/X_-$ as containing only $k$ positive /negative momentum modes respectively.   This leads to the crucial identities $n X_{+}^{-1} n=  X_{+}^{-1} n$ and  $   n X_{-}^{ -1} n = n X_{-}^{-1}$, and to the resolvent formula \footnote{See supplementary material for more details.}:
\bea \label{N}
N (\beta )=\frac{1}{\beta -\tfrac{1}{2}} 
\bigg(1- \frac{1}{\beta -\frac{1}{2}} f X_+^{-1}n X_- g \bigg)
\eea
For our application $f=g=\Theta_V$ where $\Theta_V$ is the characteristic function of $V$ ($\Theta_V(x)=1$ if $x\in V$ and is $0$ otherwise).
Eq.  \eqref{RHP} written as $\ln X (x)= \ln(X_+(x))-\ln(X_-(x))=\ln(\frac{\beta +\frac{1}{2}}{\beta -\frac{1}{2}} )\Theta _V(x)$,  has the standard solution 
\bea \label{RHPsol}
\ln X_{\pm }(x)= i h(\beta) \sum _j \ln  {e^{\frac{i}{R}(x
\mp i\epsilon )} -e^{i a_j}\over e^{\frac{i}{R}(x\mp i\epsilon )} -e ^{ib_j}},
\eea
where we introduced the shorthand notation
\bea
h(\beta 
) =\frac{1}{2\pi }\ln \frac{\beta +\tfrac12}{\beta -\tfrac 12}.
\eea
Substituting $X_{\pm}$ from \eqref{RHPsol} into (\ref{N}), we find for the resolvent:
\bea &\nonumber
\langle x|N(\beta )|y\rangle = 
\frac{1}{\beta -1/2}\delta (x-y)-\frac{e^{-
i h (Z^{+}(x)-Z^{+}(y))} }{\beta ^2-\frac{1}{4}} n(x,y) 
\eea
where 
\bea
Z^{\pm }(x)= \ln \prod _j 
(\frac{e^{ \frac{i(x\pm i\epsilon)}{R}}-e^{ \frac{i a_j}{R}}}{e^{ \frac{i(x\pm i \epsilon)}{ R}}-e^{ \frac{i b_j}{R}}} ).
\eea
This form is the finite radius generalization of the result obtained in \cite{casini2009reduced} for fermions on an infinite line.

The EH in the $\alpha \neq 0$ sector is obtained by applying (\ref{g}) to compute the resolvent and then integrating over $\beta$ as in \eqref{EH}.  Adapting the procedure described in  \cite{casini2009reduced} to our case, we find:
\bea  \label{H kernel dirac}
\!\!\!H_{V}^{\alpha\neq  0} ={4\pi^2  }e^{i\alpha \frac{(x-y)}{R}}n_{PV}(x,y)\delta (Z^+(x)-Z^+(y)),
\eea 
where $n_{PV}(x,y)=n(x,y)-{1\over 2}\delta(x-y)$ is the principal part of $n$.
Using the distribution kernel \eqref{H kernel dirac} to compute explicitly the EH, gives, owing to the $\delta$ function, a local contribution 
%
\bea     
 \!{\cal H}^{\alpha\neq 0}_{V,\text{loc.}}   \!=\!
-2\pi\!\! \int _{V}\!\! {dx}  \Psi ^{\dagger }\!\bigg[
   \frac{i}{|Z'|}\frac{d}{\text{dx}}-\frac{(1\!-\! 2 \alpha) }{2 
   R |Z'|}- \Big(\!\frac{i}{2|Z'|}\!\Big)\!' \bigg]\!\Psi, \label{H Dirac Local}
\eea
as well as a bi-local contribution with kernel:
\bea\label{bilocal dirac}
 H_{V,\text{ \text{bi-loc.}}}^{\alpha\neq 0} \!=\!  {2\pi}\!\! \sum _{l; y_l(\!x\!)\neq x} \!\!\!\frac{e^{i\alpha 
 \frac{(x-y)}{R}}|Z'\!(x)|^{-1}}{  R \big(1-e^{\frac{i(x-y)}{R}}\big)} 
\delta(x\! -\! y_l(x) ),
\eea
where $y_l(x)$ are solutions of $Z(x)=Z(y)$.

Let us pause to inspect the single interval result for $\alpha=\frac{1}{2}$.   Since the free fermion is a conformal field theory, the CFT results of \cite{myers,Wong:2013fk} suggest that the single interval EH should be an integral of the energy density $Q \propto i \Psi^{\dagger} \partial \Psi $. At first glance, \eqref{H Dirac Local} seems at odds with this form, as it contains an additional piece proportional to the number density $\rho(x) \!= \! \Psi^{\dagger}(x)\Psi(x)$. On closer inspection, we see that the CFT expectation and the direct calculation are, in fact, consistent. 
Indeed, the interpretation of the CFT local energy tensor in \cite{myers,Wong:2013fk}  as an operator, has to be done with care: in particular, the form $-i\!\int dx\, c(x)\Psi^{\dagger} \partial \Psi$, is not Hermitian by itself. However, integration by parts shows that it is made Hermitian with the addition of a term proportional to an integral over the density $(i/2)\!\int dx\, c'(x)\rho(x)$, which is exactly the extra term in Eq.  \eqref{H Dirac Local} for $\alpha=1/2$. In the path integral formalism of \cite{Wong:2013fk}, this term can be derived by identifying the EH as the generator of conformal rotations fixing the end points of $V$, which must also generate a phase rotation of the Dirac fermion due to its non-trivial conformal spin. This symmetry generator is the hermitian form of the local energy operator given by $T_{00}={-i\Psi^{{\dag}}(\partial_{x}\Psi)+i(\partial_{x}\Psi^{{\dag}})\Psi\over 2} $.  Given this definition, our result (\ref{H Dirac Local}) takes the form $ \mathcal{H_{V}}=\int_{V} dx \beta(x) T_{00}$, with  $\beta(x)\!=\!2 \pi |Z'(x)|^{-1}=4\pi R \csc{a - b\over 2 R}\sin{a- x\over 2 R} \sin{b- x\over 2 R}$  as a local entanglement temperature. Turning on a flux then introduces a shift of $T_{00}$ by a conserved charge density $-\mu \rho(x) $ with chemical potential $\mu =  \frac{(1-2 \alpha) }{2 R}$. As shown in \cite{Wong:2013fk}, this leads to a generalized first law of EE in which small excitations of the vacuum causes a change in EE given by $\delta S_{V}\!=\! \int_{V}dx\,  \beta(x)  \delta \langle (T_{00}-\mu \rho(x))\rangle$ .
 
The Ramond sector in which $\alpha \!\!=\!\!0 $ requires separate consideration, since the zero mode $k\!\!=\!\!0$ acts on a doubly degenerate ground state subspace of occupied/unoccupied modes.  We choose the state dictated by the zero-temperature limit of the Fermi-Dirac distribution. For a topological zero mode, Fermi-Dirac gives the mixed state $\tfrac 12(|\rm{occupied}\rangle \langle \rm{occupied}|+|\rm{empty}\rangle\langle \rm{empty}|)$. In this state \footnote{The following also applies to the pure state $\tfrac 12(|\rm{occupied}\rangle +|\rm{empty}\rangle$.}, the Green's function is
\bea\label {g0}
G^{\alpha=0}&\!=\!& n -\frac{1}{2}|0\rangle \langle 0| 
\eea
and our Resolvent is
\bea
L^{\alpha =0}(\beta) &\!=\! & (P_{V} n P_{V} - \frac{1}{2}+\beta -\frac{1}{2} P_{V}|0\rangle \langle 0|P_{V} )^{-1}\label{l0}
\eea
where $|0\rangle$ is the $k=0$ momentum mode $\langle x|0\rangle = \frac{1}{\surd 2 \pi  R}$.  

The difference relative to the case we have already solved is that the zero modes introduce a shift of the projector $n$ by a rank one perturbation $-\frac{1}{2}|0\rangle \langle 0|$. We proceed by treating the zero mode contribution as a Dyson perturbative expansion, which we subsequently re-sum (see supp. material), leading to
\begin{equation}
\label{L0}
L^{\alpha=0}(\beta)=N(\beta)+ \frac{N(\beta)P_V|0\rangle \langle 0|P_VN(\beta)}{2- \langle 0|P_V N(\beta)P_V|0\rangle }.
\end{equation}
We note in passing that  Eq. (\ref{L0}) may be applied for computing the EH for single particle excitations of the vacuum and that the re-summation also works for higher rank perturbations.

While the first term on r.h.s of Eq. (\ref{L0}) will yield a similar Hamiltonian as we considered up to now, the second term on the r.h.s. is different: it is due to the R sector zero mode and produces a non-local contribution to the EH.   Explicitly
\begin{eqnarray}&
\langle x|L^{\rm R}_{\text{zero-mode}}(\beta )|y\rangle \!=\! \frac{2 \sinh^2(\pi  h)  e^{i h( Z(y)-Z(x))}}{  \pi R  (1
\!+\! e^{ {{l}_{v} h\over R}})}, 
\end{eqnarray}
where $l_{v}=\sum_{i}(b_{i}-a_{i})$ is the total length of $V$. Using \eqref{EH},
the contribution from $L^{\rm R}_{\text{zero-mode}}$ to the EH is:
\begin{eqnarray}\label{HzeroMode}
&H_{V\,\text{zero-mode}}^{\rm R}
\!=\! 
{-1\over R}  \underset{- \infty }{\overset{\infty }{\int }} d h  
\frac{1}{1+e^{\frac{l_{v} h}{R}}}e^{i h( Z(y)-Z(x))}
\\ &=
\sum_ {l}{- \pi\over |Z'(x)|R}\delta(x-y_{l}(x))+p.v.\frac{\pi i}{2 l_{v}} {1\over \sinh\big ({\pi R\over l_{v}}( Z(y)-Z(x))\big)} \nonumber
\end{eqnarray}
{\it Thus, the zero mode induced part of the EH has a non-local contribution, even for a single interval.}  
It is remarkable that for one interval the $\delta(x-y)$ term exactly cancels the ${1/2}$ shift in $\alpha$ (the "chemical potential" term) in  \eqref{H Dirac Local}, and we get that, as far as the {\it strictly local} terms are concerned, $H_{V,\text{loc.}}(\alpha=0)=H_{V,\text{loc.}} (\alpha=1/2)$.

The non-local contributions to the resolvent due to the zero mode also changes the EE in the R sector relative to the NS. 
The EE $S_V=-{\rm Tr} \rho_V\ln \rho_V$ can be expressed as in integral form:
\bea
\label{S integral}&\!\!\!\!\!\!\!\!
S_{V}\equiv  \! \int _{{1\over 2}}^{\infty}\!{\rm d}\beta \text{Tr}[ (\beta -1/2)(L (\beta)\!-\! L (-\beta))\!-\!\frac{2\beta }{\beta +1/2}]
\eea

It follows, using \eqref{Lalpha}, that all Dirac fermions with $\alpha\neq 0$ have the same EE (disregarding possible anomaly contributions coming from the UV cutoff). 

Using \eqref{S integral} we find the contribution to the EE from the change in BC, $\delta  S\equiv S_{R}-S_{NS}$: 
\bea &
\!\!\!\delta S ={l_{v} \over 2 R }\int_0^{\infty}dh \,
{\rm tanh}(\frac{l_{v} h}{2R})  (\coth(h \pi )-1). \label{deltasint}
\eea
Expanding in the ratio 
$\frac{l_v}{R}$ gives:
\bea
\delta S \!\sim\! \sum_{n=1}^{\infty }
   \frac{\left(2^{2n}\!-\!1\right)\! B_{2n} \zeta(2n)l_{v}^{2n}}{2n (2\pi R)^{2n}}\!\sim\! \frac{l_{v}^2}{48 R^2}-\frac{l_{v}^4}{ 5760  R^{4} } \label{deltas},
\eea
where $B_{2n}$ are Bernoulli numbers. 
Remarkably, using the replica trick, the authors of \cite{herzog2013entanglement}
have given the change in the EE induced by BC change for a massless Dirac fermion in the form of the series (\ref{deltas}), and noted that the series is not convergent. Here we have obtained the exact expression Eq. (\ref{deltasint}) for the EE and can now identify the replica trick result of \cite{herzog2013entanglement} as an asymptotic expansion of the convergent integral (\ref{deltasint}).
Finally, we consider the limiting case where $V$ is the entire circle. Plugging $l_{v}\rightarrow 2\pi R$  into \eqref{deltasint}, we find $\delta S=\log 2$, as expected from the degeneracy \footnote{The requirement of Wick's theorem to hold in the presence of zero modes, forces the state on the full circle to be a mixed state, hence the nonzero entropy for the full circle in the R sector.}. 

Finally, we find that a similar story holds for the Majorana fermion. We derived a similar relation between the Green's function $G_{\cal M}(x,y)=\langle\psi(x)\psi(y)\rangle$ and EH \footnote{Supplementary Online Material.}: 
\bea
H_{V}^{{\cal M}}=  \frac{1}{2} \ln(  (P_V G_{\cal M} P_V ){}^{-1}-1)
\eea 
This relation differs from the Dirac one by the important factor of $-1/2$, while $G_{\cal M}(x,y)$ agrees with the Dirac case in the NS and R sector.   In addition, since $\psi(x)^{2}$ is a constant for Majoranas, we can ignore the $\d (x-y)$ terms in the EH kernel. Thus one finds that in the NS sector, the Majorana EH has the local term 
\bea
{\cal H}_{V, \text{loc.}}^{\rm{NS}}\!=\! \pi  \!\! \int_{V} \!\!  dx  {i\over Z'(x)}
 \psi (x)\partial _x\psi  (x)
\eea
while the bilocal kernel (\ref{bilocal dirac}) remains the same.    
In the R sector the situation is more subtle, as the Majorana zero mode $b_0$ has no natural partner to create a complex fermion.   To get the minimal, non trivial Hilbert space representation of the Clifford algebra containing $b_0$, we have to introduce an additional anti-commuting operator.     A physical realization of such a situation is the edge of a $p+i p$ superconductor, in the presence of a ($\pi $ flux) vortex in the bulk  (see, e.g. \cite{fendley2007edge}). Such a vortex acts to change the BC on the edge into R type and binds a Majorana zero energy mode to its core. We can combine the Majorana operator $b_0'$  for the bulk-core zero mode with our edge zero mode $b_0$, forming a complex/Dirac fermion operator ground state \(a_0=\frac{1}{\surd2} (b_0+ i b_0' )\) which switches between it's associated ``$|\rm{occupied}\rangle$'' and ``$|\rm{empty}\rangle$'' ground states.  Since $b_{0}^{2} =\frac{1}{2}$ by the Majorana anti-commutation relations, the Green's function is same as the Dirac one in (\ref{g0}) for any choice of ground state.  Up to the $-\frac{1}{2}$ factor, the resulting R sector single particle EH is equal to the Dirac one, while the EE changes by a factor of $\frac{1}{2}$.

To summarize, in this letter we provide a framework for the computation of EH in the presence of topological defects by combining RHP and finite rank perturbation theory. We present the first exact calculation of an EH on a compact space, with and without zero modes. We find that for a single interval the Hamiltonian consists of a local term in agreement with the CFT prediction for a single interval, as well as a non-local term associated with the zero mode and with the possibility of several entangling intervals. 
Note that our method also applies for an exact, but non-topological zero mode, however, any small perturbation will shift such a mode away from exact zero and the ground state will be unique as opposed to the mixed state we are considering. 
We also present an expression, Eq. \eqref{deltasint}, 
 representing the {\it local} entropic signature of the presence of a boundary changing operator that reproduces the boundary entropies computed previously.
Immediate possible applications for the method are access to the EH and EE of excited states, systems with non-chiral fermions and systems with several fermion flavors. While we concentrated here on the simple case of a scalar RHP, a full matrix problem appears immediately in more involved situations, in particular, for non-relativistic fermions with a finite Fermi sea \footnote{G. Wong, I. Klich and D. Vaman, {\it in preparation}.}. 

{\it Acknowledgement:} The work of IK was supported by the NSF CAREER grant DMR-0956053. DV and GW were supported in part by the U.S. Department of Energy under Grant No. DE-SC0007984. GW would like to thank Leo Pando Zayas and the Michigan Center for Theoretical Physics for hospitality during the initial stages of this project, Jean Marie Stephan, Jihnho Baik, Jeffrey Teo for helpful discussions.  GW was supported in part by the UVA presidential fellowship.

\section{Supplementary Material}
\subsection{ Entanglement Hamiltonian for Majorana Fermions} 
Consider a chiral Majorana fermion in $(1+1)$d  with Lagrangian $\tfrac{ i}2  \psi (\partial_t+\partial_x)\psi $, where the spatial dimension is periodic $x\sim x+2\pi R$. 
The canonical equal-time anti-commutation relation is $\{
\psi(x),\psi(y)\}\!=\!\delta(x-y)$,  and we require $\psi^{\dag}(x)\!=\!\psi(x)$. Given  the 
mode expansion \bea \psi (t,x)=\frac{1}{\sqrt{2\pi  R}} \sum _k b_k e^{-i\tfrac{k}R (t-x) }, \eea 
with $k\in \mathbb{Z}$ for the R sector and $k\in  \mathbb{Z}+\frac{1}{2}$ for the NS sector, the ground state satisfies $b_k|\Omega \rangle  =0$  for $k >0$.  As a single particle operator, the equal-time Green's function in the NS sector is essentially the same as in the Dirac case, i.e.:
\bea 
G^{\rm NS}(x,y)\!\equiv\! \langle \Omega|\psi  (x+ i 0^+  )\psi (y)|\Omega\rangle\equiv e^{\frac{i  (x-y)}{ 2R}}n(x,y),
\eea
 where $n(x,y)\equiv\langle x|n|y\rangle$ denotes the same projector as in the Dirac case.   

Given the reduced density matrix $\rho_V$ on a region $V$, the single particle EH for Majorana fermions, $H_{V}^{{\cal M}}$,  is defined by:
\bea & \rho_V={\cal Z}_V^{-1}e^{-{\cal H}_V^{\cal M}}~;\\ & {\cal H}_V^{{\cal M}}=\int_{ V}d x\,d y H_{V}^{{\cal M}}(x,y)\psi(x)\psi(y).\eea 
We now show that that $H_V^{\cal M}$ is obtained from the Majorana Green's function by: 
\begin{equation} \label{MEH}
H_{V}^{{\cal M}}=  \frac{1}{2} \ln(  (P_V G P_V ){}^{-1}-1),
\end{equation}
where $P_V$ is a projector onto $V$, with 
$ \langle x |P_V |y \rangle = \Theta _V(x) \delta(x-y)$. Here $\Theta_V$ is the characteristic function for $V \in S^{1}$, with $\Theta _V(x)=1$ if $x\in V$ and $\Theta _V(x)=0$ otherwise.  We note that EE has been studied extensively also in terms of Majorana fermions (see e.g. \cite{vidal2003entanglement}), but are not aware of a simple treatment of the EH as we present below. 
To derive Eq. (\ref{MEH}), we take the Majorana operators to be defined on a discrete lattice, with $\{\psi_i,\psi_j\}=\delta_{ij}$, and compute the correlation functions from the trace formula for paired states in \cite{klich2014note} adapted to the present notation:
\begin{equation}
{\rm Tr} e^{A_{ij}\psi_i\psi_j}e^{B_{kl}\psi_k\psi_l}=\sqrt{det(1+e^{2A}e^{2B})}
\end{equation}
where $A,B$ are antisymmetric matrices.
We have:
\bea&
G_{ij}\equiv\bra \psi_i \psi_j\ket
= {1\over 2}\big(\delta_{ij}+
\partial_\alpha \log {\rm Tr} e^{-  \cal{H}_V^{\cal M}}e^{{\alpha}  (\psi_i \psi_j-\psi_{j}\psi_{i})}|_{\alpha=0}\big)  \nonumber\\  & ={1\over 2}\delta_{ij}+
{1\over 2}\partial_\alpha \log \sqrt{\det (1+e^{-2 H_{V}^{{\cal M}}}e^{{2\alpha}(|i\rangle\langle j|-|j\rangle\langle i |) })}|_{\alpha=0}\nonumber\\  & = \Big(\frac{1}{1+e^{-2 H_{V}^{{\cal M}}}}\Big)_{ij},
\eea
where the antisymmetry of $H_{V}^{\cal M}$ was used in going to the last line.
Inverting these relations gives \eqref{MEH}. 

\subsection{Green's function in the Ramond sector}
In evaluating the Green's function in the R sector, we are allowed to choose any linear combination of the ground states, or their mixed state. However, in deriving the EH, we assume states that are uniquely described by the Green's function through Wick's theorem. Such states must have a vanishing expectation values for products of an odd number of Majorana operators, and restrict our choice.  

This is ensured by taking states that preserve fermion parity. Fermion Parity $(-1)^F$ acts by $(-1)^F \psi (-1)^F \!=\! - \psi $. For states $\rho $ that commute with $(-1)^F$, we have: 
\bea & -Tr(\rho \psi )=Tr ( \rho (-1)^F \psi (-1)^F ) = Tr(\rho \psi )  \\ \nonumber & \Rightarrow Tr(\rho \psi ) =0.
\eea
In the R sector, taking into account $b_0$, we find
\bea
G^{\rm R}\!(x,y)
\!=\!\frac{1}{2 \pi
 R} (\!-\!\frac{1}{2}+\sum _{k=0}^{\infty }  e^{\frac{i k (x-y+i0^+)}{R} })= G^{\alpha=0}(x,y).
\eea 
\subsection{The relation between the resolvent and the Riemann Hilber problem solution.}
Here we check that the expression \eqref{N} is indeed the inverse of $K$.  For this, one checks that 
\bea fng-fX_+^{-1}nX_-g-{1\over \beta-1/2} fngfX_+^{-1}nX_-g=0,\eea
 where for simplicity we restricted to the case of $i=1$. 
Here $f,g, X_\pm$ are understood to be operators which act multiplicatively in position space, with $f g = (\beta-1/2)( X_-^{-1}X_+-1)$. Substituting into the last term in the previous expression leads to 
$f(n - X_+^{-1} n X_--n(X_-^{-1} X_+ -1)X_+^{-1}n X_-)g=0$.  This equality can be checked using the identities \bea & X_{+}^{-1} n=n X_{+}^{-1} n , \\ & n X_{-}^{-1}=n X_{-}^{ -1} n ,\eea  which follow from the fact, that by their definitions, $X_{+} $ contains only positive $k$s and therefore multiplication by $X_{+}$ can only increases momentum while, similarly, multiplication by $X_-$ can only decrease momentum.

\subsection{Rank one perturbation and resummation}
Consider a matrix $A$ and a rank one perturbation $B\!=\! |s\rangle \langle s|$ we can formally expand in  $A^{-1}B$ and then re-sum: 
\bea & 
(A+B)^{-1} = (1+A^{-1}B )^{-1} A^{-1}   \\ \nonumber
&= A^{-1} - A^{-1}|s\rangle (\sum_{n=0} (-1)^{n} \langle s | A^{-1} | s \rangle^{n})  \langle s | A^{-1} \\  \nonumber & = A^{-1} - \frac{A^{-1}|s\rangle\langle s | A^{-1}}{1+ \langle s | A^{-1} | s \rangle} \label{resummation}
\eea
Substituting $A=P_{V} n P_{V} - 1/2 +\beta$ and $ |s\rangle = \frac{i}{\sqrt{2}} P_{V}|0\rangle $ then gives the desired resolvent
\begin{equation}
L^{\alpha=0}(\beta)=N(\beta)+ \frac{N(\beta)P_V|0\rangle \langle 0|P_VN(\beta)}{2- \langle 0|P_V N(\beta)P_V|0\rangle }.
\end{equation}
Written explicitly using the kernels of the operators above the resolvent kernel is given by:
\bea\nonumber
\langle x|L^{\rm R}_{\text{zero-mode}}(\beta )|y\rangle \!=\! \frac{1}{2 \pi  R}\frac{\int_V dz dz' \langle x|N(\beta )|z\rangle  \langle z'|N(\beta )|y\rangle }{2- \frac{1}{2 \pi
 R}\int_V dzdz'\langle z|N(\beta )|z'\rangle }.
\eea

\end{document}